\documentclass[letterpaper,english]{extarticle}
\usepackage[T1]{fontenc}
\usepackage[latin9]{inputenc}
\usepackage{textcomp}
\usepackage{amsmath}
\usepackage{amssymb}
\usepackage{graphicx}
\usepackage[numbers]{natbib}

\makeatletter

\newcommand{\lyxrightaddress}[1]{
	\par {\raggedleft \begin{tabular}{l}\ignorespaces
	#1
	\end{tabular}
	\vspace{1.4em}
	\par}
}

\makeatother

\usepackage{babel}
\begin{document}
\title{Supermassive Black Holes from Bose-Einstein Condensed Dark Matter\\
- or Black and Dark Separation by Angular Momentum - }
\author{Masahiro Morikawa}
\maketitle

\lyxrightaddress{Department of Physics, Ochanomizu University \\
2-1-1 Otsuka, Bunkyo, Tokyo 112-8610, Japan\\
hiro@phys.ocha.ac.jp }
\begin{abstract}
Many supermassive black holes (SMBH) of mass $10^{6\sim9}M_{\odot}$
are observed at the center of each galaxy even in the high redshift
($z\approx7$) Universe. To explain the early formation and the common
existence of SMBH, we proposed previously the SMBH formation scenario
by the gravitational collapse of the coherent dark matter (DM) composed
from the Bose-Einstein Condensed (BEC) objects. A difficult problem
in this scenario is the inevitable angular momentum which prevents
the collapse of BEC. To overcome this difficulty, in this paper, we
consider the very early Universe when the BEC-DM acquires its proper
angular momentum by the tidal torque mechanism. The balance of the
density evolution and the acquisition of the angular momentum determines
the mass of the SMBH as well as the mass ratio of BH and the surrounding
dark halo (DH). This ratio turns out to be $M_{BH}/M_{DH}\approx10^{-3\sim-5}(M_{tot}/10^{12}\mathrm{M}_{\odot})^{-1/2}$
assuming simple density profiles of the initial DM cloud. This estimate
turns out to be consistent with the observations at $z\approx0$ and
$z\approx6$, although the data scatter is large. Thus the angular
momentum determines the separation of black and dark, \textsl{i.e.
}SMBH and DH, in the original DM cloud. 
\end{abstract}

\section{\label{sec:level1}Introduction}

Recent observations have revealed a tremendous number of supermassive
black holes (SMBH) of mass $10^{6}-10^{10}M_{\odot}$ in the wide
range of redshift up to $z\approx7.642$\citep{Wang2021,Matsuoka2018,Trakhtenbrot2019}.
SMBH seems to be a common astronomical species everywhere in the Universe
from the very early stage. Most of the galaxies have SMBH in their
center as if the SMBH even defines the center of the galaxy. Furthermore,
several universal correlations of SMBH with the galactic properties
are widely reported so far\citep{Gultekin2009}. Thus, SMBH must be
a crucial component in the galaxies and deeply connected with their
evolution. The remarkable properties of SMBH are their very early
formation and their common existence associated with galaxies: both
of them are still not yet well understood. We would like to consider
the fundamental origin of such SMBH. 

It will be natural to consider that SMBH may be formed from massive
primordial stars or directly from the gas clouds\citep{Rees1978,Rees1984}.
For example, the model of a direct collapse of a gas cloud assuming
subsequent steady gas accretion with the Eddington limiting rate can
explain the early formation of huge SMBH\citep{Loeb1994,Woods2018}.
However, such ideal conditions, strong ambient radiation, low metallicity,
steady high accretion,...\citep{Yoshida2019}, are considered to be
rare and therefore the early stage SMBH should not be common. However,
it may happen in the future that much more SMBHs are commonly observed
in the early Universe, then we need to consider any universal formation
mechanisms of them. Furthermore, we may need to explore a general
model of SMBH formation for explaining the universal association and
correlation of SMBH with galaxies. 

Thus, we have developed an alternative model of SMBH formation from
the coherent Bose-Einstein Condensed (BEC) matter which is supposed
to compose the dark matter (DM)\citep{Morikawa2016,Morikawa2019}.
This BEC-DM model belongs to the category of the scalar-field DM model
families studied for many years \citep{Baldeschi1983,Membrado1989,Sahni2000}.
BEC-DM model is now actively studied which is partially summarized
in the references in \citep{Cr=000103ciun2020}. Since BEC is coherent
and has almost no velocity dispersion, it can naturally collapse toward
SMBH very fast in the isolated ideal cases. This model of BEC-DM arises
from a simple idea that DM and dark energy (DE) are the same boson
field but their phases are different for DM and DE\citep{Nishiyama2004}.
In the original model, the uniform coherent condensation of some low-mass
boson is supposed to form global DE, which causes the exponential
cosmic expansion. On the other hand, the non-uniform condensation
and the boson in the normal gas-phase form local DM. Then, it will
be natural to consider that the locally coherent component BEC collapses
into black holes of arbitrary mass. Thus, all the dark and black species
DE, DM, and SMBH are considered to be, or originated from, the same
boson field in this mode, but only their phases are different from
each other.

However, in the actual Universe, we encountered a serious problem
of the angular momentum which strongly prevents the SMBH formation
even for the coherent BEC-DM. Previously in \citep{Morikawa2019},
we have assumed a special DM Axion which has a sufficiently small
mass and is supposed to form BEC. In this case, the trigonometric
potential of the Axion yields a tiny attractive self-interaction force.
Even if it were small, this attractive self-interaction can dominate
the angular momentum barrier and allow the BEC-DM to collapse to form
SMBH. In this case, the Axion mass parameter should have an appropriate
value for the successful description for the observed ratio of DM
and dark halo (DH) surrounding the galaxy. It may be favorable if
there were more general and common mechanism to avoid this angular
momentum barrier, because SMBH is considered to be a common structure
in our Universe.

Therefore, in this article, we consider another approach to solve
the angular momentum problem without assuming undiscovered fields
such as Axion. We do not attempt to overcome the angular momentum
barrier as in the case of the Axion model, but we focus on a special
epoch appropriate for SMBH formation. We consider the very early stage
of the Universe when each DM mass clump was just acquiring first angular
momentum by the tidal torque effect\citep{Peebles1969,Sugerman2000}.
The increase of the angular momentum tends to prevent the SMBH formation
but the increase of the mass enhancement promotes the SMBH formation.
Then the balance of these two effects should determine the mass of
the SMBH as a function of the clump mass.

In this context, the angular momentum does not prevent the structure
formation but adjusts the size of the structures. In our context,
the angular momentum adjusts the separation of the central SMBH and
the surrounding DH, both of them are essential for the galaxy formation.
Furthermore, the angular momentum guarantees the stability of astronomical
objects as well. This can be clarified if we estimate the angular
momentum in various mass scales in the Universe. Then we will find
the scaling relation $J\approx10^{4}GM^{2}/c$ over 30 digits of objects
composed from dust, Hydrogen gas, and DM. This further motivates the
formation of SMBH from DM.

We first consider the present cosmic angular momentum on various scales
in section 2. The angular momentum in the wide range of scales provides
us with a rough estimate of time scales for the formation of BH and
other compact structures. Then in section 3, we study the SMBH formation
from BEC-DM. Effective potential and the BH formation conditions are
discussed in the non-relativistic approximation. In section 4, we
consider how the SMBH and DH are separated by the balance of the angular
momentum acquisition and the density enhancement in the early stage
of the Universe. In section 5, we describe the verification of our
analysis comparing with the observational data. In the last section
6, we summarize our argument showing further developments for future
work. 

\section{Angular momentum }

\begin{figure}
\includegraphics[width=12cm]{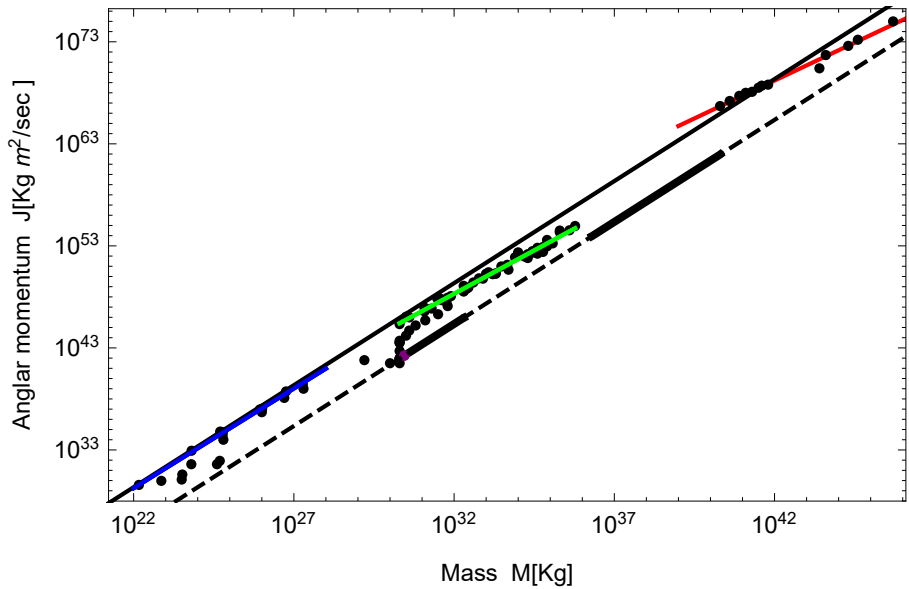}\caption{The angular momentum of various astronomical objects is shown\citep{deVaucouleurs1970,Larson1979,Muradian1999}.
They roughly scale as $J=\kappa\frac{G}{c}M^{2},\quad\kappa\approx10^{4}$(black
solid line) on the whole. According to this figure, astronomical objects
are divided into three classes: dust(metal), gas(non-metal), and DM,
from small to large scales. Each of them locally scales as $J\propto M^{\gamma}$
where $\gamma\approx1.95,1.70,1.48$, respectively represented as
blue, green, and red lines. \protect \\
\label{fig:The-angular-momentum}\protect \\
}
\end{figure}

All the structure in the present Universe is rotating. They may collide
with each other and the entire rotation may not be apparent. Even
those cases, objects inherit the angular momentum to smaller or larger
scales from the original scale. The angular momentum $J$ associated
with astronomical objects in the Universe shows a conspicuous scaling
relation in the wide range of mass scales $M$ from the planets to
the galaxy clusters. This scaling is shown in Fig.\ref{fig:The-angular-momentum}
by the black solid line, 
\begin{equation}
J=\kappa\frac{G}{c}M^{2},\label{eq:1}
\end{equation}
where $\kappa\approx10^{4}$ is a constant. All the dots represent
observational data for individual astronomical objects. In the figure,
the dashed line represents the critical angular momentum, Eq.(\ref{eq:1})
with $\kappa=1,$ beyond which the black holes cannot exist. On this
dashed line, the observed black hole ranges are marked by thick solid
lines. 

According to Fig.(\ref{fig:The-angular-momentum}), the observed data
points seem to separate into three classes, structures made from dust,
gas, and DM, from small to large scales. The structures made from
dust, classified as \textit{metal} in astronomy, are rocky planets
and smaller substructures such as satellites (class 1). The structures
made from the gas of Hydrogen and Helium, classified as \textit{non-metal}
in astronomy, are the interstellar gas and the stars (class 2). The
structures made from DM, as well as a small fraction of Baryon, are
the galaxies and their clusters (class 3). In between these three
classes, there seem to be two ranges of BH: stellar scale BH ($1-10^{2}M_{\odot}$)
and SMBH ($10^{6}-10^{10}M_{\odot}$).

These scaling relations may be produced by simple physics\citep{Nakamichi2009}.
We shall regard the objects, dust, gas, and DM, like fluids. Then
the turbulence of those fluids, as well as self-gravity, may yield
the scaling relation. The Kolmogorov model of turbulence \citep{Kolmogorov1941}
describes the hierarchical steady flow of energy, in the wavenumber
space, from large to small scales, and at the smallest scale, the
energy dissipates into heat. The only relevant parameter in this picture
is the dissipation speed at the smallest scale, or equivalently, the
steady energy flow per unit mass $\epsilon$ in the whole scaling
range of the fluid.

In the self-gravitating case in the Universe, the energy flow may
be steady but the direction would be small to large scales, opposite
to the usual three-dimensional fluid case. However, the material flows
from large toward small scale, and at the smallest scale, they sediment
to form compact objects. This design of hierarchy yields the typical
relative velocity $\sigma$, at two positions separated by $r$, as
\begin{equation}
\sigma=(r\epsilon)^{1/3}.\label{eq:vel-turbulence}
\end{equation}
where $\epsilon$ represents the energy flow per mass in the wavenumber
space. Further, in the present cosmic case, a self-gravitating system
of mass $M$ and size $r$ in the virial equilibrium is characterized
by the relation
\begin{equation}
\sigma^{2}=\frac{GM^{2}}{r},\label{eq:vel-SGS}
\end{equation}
where $G$ is the gravitational constant. Combining these Eqs.(\ref{eq:vel-turbulence},\ref{eq:vel-SGS}),
we can estimate the mass of the object as \citep{Nakamichi2009}
\begin{equation}
M=G^{-1}\epsilon^{2/3}r^{5/3},\label{eq:M-scaling}
\end{equation}
and the angular momentum as
\begin{equation}
J=\frac{2}{5}G^{4/5}\epsilon^{-1/5}M^{9/5}.\label{eq:J-scaling}
\end{equation}
This Eq.(\ref{eq:J-scaling}) yields the power index $9/5=1.8$ in
Fig.\ref{fig:The-angular-momentum}, although the actual indices are
slightly different from each other: 1.95 for class 1, 1.70 for class
2, and 1.48 for class 3, and the overall slope is 2. These slightly
different indices may reflect a) the huge compressible nature of the
fluids, and b) the effect of the prevailing magnetic field which locally
enhances the flow of energy and angular momentum. Note that Eq.(\ref{eq:vel-turbulence})
can be obtained only for the non-compressible fluid. 

If we regard that the above scaling law reflects the hierarchical
turbulent structure in each class of ingredient, the parameter $\epsilon$
determines the steady flow of energy, angular momentum, as well as
the ingredient object. However, in the self-gravitating system, what
the parameter $\epsilon$ determines is not literal `dissipation`,
but the sedimentation of the material which finally forms compact
structures at the smallest scale in each class. All the compact structures
in the Universe may be formed in this flow as stagnation mainly at
the smallest scales in each class.

According to this picture, we may naturally expect that the dust forms
planets, the gas forms stars as well as stellar size BH, and in particular,
the DM forms galaxies as well as SMBH. Thus, this hierarchical turbulent
picture naturally motivates us to examine the scenario of SMBH formation
from DM\citep{Nakamichi2009}. However, Fig.\ref{fig:The-angular-momentum}
tells us more. In the case of class 2, the route from gas to the stellar
size BH seems apparent: large-scale interstellar gas continuously
connects down to the stars and BHs as the data points show. On the
other hand, in the case of class 3, the route from DM to the SMBH
cannot be observed. This discontinuity may indicate that the SMBH
cannot be formed in the steady process of the hierarchical turbulent
scenario. This is another motivation for us to consider the dynamical
stage of galaxy formation in the early Universe. Let us continue the
argument of the hierarchical turbulent scenario for a while. 

The standard Kolmogorov dissipation time scale would correspond to
the structure formation time scale in our cosmic case. The ordinary
turbulent fluid dissipates its energy into heat mainly at the low-end
of the Kolmogorov scaling. In the cosmic case, since the gravity confines
such drop-out material into compact objects, the structure formation
must be taking place at each smallest end of the class. This stagnation
in the steady turbulent flow may be the general origin of astronomical
structures such as planets, stars, and galaxies. 

\begin{figure}
\includegraphics[width=12cm]{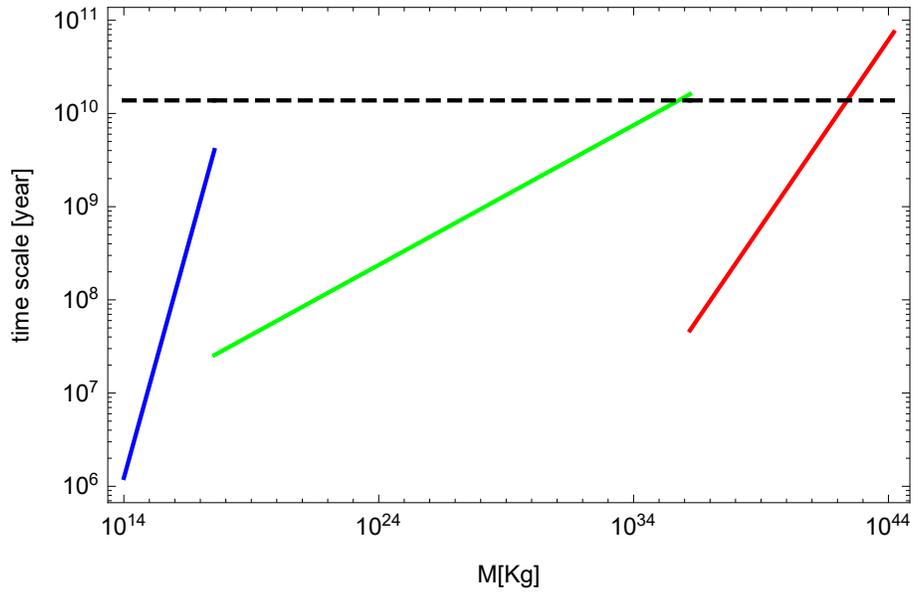}\caption{Time scales estimated from Eqs.(\ref{eq:timescales},\ref{eq:rhod}-\ref{eq:epsilons})
for each class are shown as a function of the mass $M$ of the astronomical
object. The color blue, green, and red reflects each class and corresponds
to the fitting color in Fig.(\ref{fig:The-angular-momentum}). Although
the structure formation will dominantly take place at the smallest
scale in each class, there may take place some formation process at
the intermediate scales in the steady flow of energy. The estimate
in this figure should not be taken too seriously since the Universe
is not steady, in particular at larger scales. Actually, the estimated
time scale well exceeds the age of the Universe in those scales. \protect \\
\label{fig:Time-scales}\protect \\
}
\end{figure}

We now estimate the time scales of such compact structure formations.
In each class, the actual mass density scaling is different. Best-fitting
the observed densities of astronomical objects in each class, the
mass densities are given by (\citep{deVaucouleurs1970,Larson1979,Muradian1999}),
in the unit $\mathrm{Kg}/\mathrm{Meter^{3}}$,
\begin{align}
\rho_{d} & =6.8\times10^{5}\left(\frac{r}{\mathrm{Meter}}\right)^{-0.35},\label{eq:rhod}\\
\rho_{g} & =6.3\times10^{16}\left(\frac{r}{\mathrm{Meter}}\right)^{-1.75},\label{eq:rhog}\\
\rho_{D} & =1.6\times10^{9}\left(\frac{r}{\mathrm{Meter}}\right)^{-1.33},\label{eq:rhoD}
\end{align}
for each class: dust, gas, and DM, respectively. If we adopt these
phenomenology, then the best fit of the angular momentum $J$ in Fig.\ref{fig:The-angular-momentum}
yields the approximate Kolmogorov parameters as, 
\begin{equation}
\epsilon_{d}=10^{-4}\epsilon_{D},\;\epsilon_{g}=10^{-2}\epsilon_{D},\;\epsilon_{D}=3.0\times10^{-5},\label{eq:epsilons}
\end{equation}
in the unit $\mathrm{Meter^{2}Sec^{-3}}$. They determine the steady
energy flow, from small scale to larger scale, per mass per time.
Thus, it may be possible to estimate the time scales for the formation
in each scale. During the formation time $\tau$, the total amount
of energy flow out is the gravitational potential of the compact object
$GM^{2}/r$, and the flow speed per mass is $\epsilon$. Therefore
the formation time scale $\tau$ of this compact object is estimated
as 
\begin{equation}
\tau=G\rho r^{2}\epsilon^{-1}.\label{eq:timescales}
\end{equation}
This estimated time scales are expressed in Fig.\ref{fig:Time-scales}. 

According to Fig.\ref{fig:Time-scales}, the formation time scale
of the object of mass $10^{9}M_{\odot}\approx2\times10^{39}$Kg reads
about $10^{9}$ years, corresponding to the redshift $4.8$, which
is marginal or ruled out by observations. Further, the large-scale
region of this diagram will not be appropriate since the estimated
time scale is based on the steady process for structure formation.
Since the Universe evolves from the initial condition and therefore
the structures are not at all steady. Moreover, the right end of the
time scale exceeds the age of the universe. We need to consider the
dynamical origin of SMBH going back to much earlier stages in the
Universe. We will discuss this issue in the subsequent sections.

\section{BEC-DM to SMBH }

The standard DM in the gas phase induces velocity dispersion and cannot
easily collapse to a BH. On the other hand, if the bosonic DM were
in the quantum condensed phase, or in Bose-Einstein condensed (BEC)
phase, it behaves as a classical scalar field without dispersion.
Thus DM in the form of BEC can collapse to form a BH. We consider
this possibility in this section\citep{Morikawa2019}.

BEC becomes possible when the wave functions of atoms overlap with
each other, or the thermal de-Broglie length $\lambda_{dB}\equiv\left(2\pi\hbar^{2}/(mkT)\right)^{1/2}$
exceeds the mean separation of atoms $r\equiv n_{}^{-1/3}$, where
$n$ is the boson number density, $m$ is the boson mass, and $k$
is the Boltzmann constant\citep{Castin2001,Courteille2001}. This
gives the BEC condition and defines the critical temperature $T_{c}\equiv2\pi\hbar^{2}n^{2/3}/(mk)$.
This temperature dependence on $n$ is the same as the cosmic number
density $T\propto n^{2/3}$. Therefore, once the cosmic gas cools
below the BEC critical temperature, it keeps the BEC phase all later
times\citep{Nishiyama2004}. This guarantees the robustness of the
BEC phase in the Universe although the BEC phase will be destroyed
by non-adiabatic processes such as violent collisions.

The BEC is the macroscopic coherent wave and is described by the classical
wave $\psi\left({t,{\bf {x}}}\right)$ which evolves as 

\begin{equation}
i\hbar\frac{\partial\psi\left(t,{\bf x}\right)}{\partial t}=\left(-\frac{\hbar^{2}}{2m}\Delta+m\phi+g\left|\psi\right|^{2}\right)\psi,\label{eq:GPE}
\end{equation}
where $\phi\left({t,{\bf {x}}}\right)$ is the gravitational potential
field, and generated by BEC,
\begin{equation}
\Delta\phi=4\pi Gm\left|\psi\right|^{2}.\label{eq:Poisson}
\end{equation}
These are the non-relativistic approximation of BEC and are widely
studied for describing DM for many years\citep{Sin1994,B=0000A8ohmer2007,Fukuyama2008,Gupta2017}. 

Within this non-relativistic approach, the BH formation will be described
by the condition that the portion of DM enters inside the corresponding
Schwarzschild radius. However, there is another necessary condition
for BH formation for the boson case. For the boson system, the structure
is supported by the quantum pressure against its gravity. A BH is
formed when this balance is no longer sustained, \textit{i.e.} when
the Schwarzschild radius $r_{s}=2GM/c^{2}$ exceeds the boson Compton
wavelength $\lambda_{c}=2\pi\hbar/(mc)$, ($r_{S}>\lambda_{c}$),
\begin{equation}
M>M_{K}\approx\pi\frac{m_{P}^{2}}{m},\label{eq:Kaup}
\end{equation}
where $m_{P}=(c\hbar/G)^{1/2}$is the Planck mass. This limiting mass
$M_{K}$ is called the Kaup mass, and the exact value $M_{K}=0.633m_{P}^{2}/m$
is derived from numerical calculations based on general relativity\citep{Kaup1968}.
This mass is far smaller than the popular limiting mass for Baryons
$m_{P}^{3}/m^{2}$. However, this Kaup mass sets the lower bound of
the BH formed from bosonic DM. For example, if the boson mass is $10^{-16}\mathrm{eV}$,
then the BH of mass lower than $0.84\times10^{6}M_{\odot}$ cannot
be formed. 

The above set of non-linear equations (\ref{eq:GPE}-\ref{eq:Poisson})
is still difficult to solve and to get general insight, although numerical
calculations themselves would be relatively easy. In order to be analytical
as much as possible, let us assume that the self-interaction is very
small and that the anisotropic back reaction from BEC field to gravity
is negligible. These approximations will be justified since we do
not rely upon the self-interaction and the acquired angular momentum
turns out to be very small in our subsequent calculations. Then, Eq.(\ref{eq:GPE})
allows the separation of variables, 

\begin{equation}
\psi\left(t,\boldsymbol{x}\right)=\varphi(t,r)Y_{l}^{m}(\theta,\varphi),\label{eqvarsep}
\end{equation}
where $Y_{l}^{m}(\theta,\varphi)$ is the spherical harmonics. Now,
the Lagrangian for the fields $\varphi(t,r)$ and $\phi\left(t,r\right)$
becomes, 

\begin{equation}
\begin{array}{cc}
L= & \frac{\text{i\ensuremath{\hbar}}}{2}(\varphi^{\dagger}\dot{\varphi}-\dot{\varphi}^{\dagger}\varphi)-\frac{\text{\ensuremath{\hbar}}^{2}}{2m}(\partial\varphi^{\dagger}/\partial r)(\partial\varphi/\partial r)-\frac{g}{2}\left(\varphi^{\dagger}\varphi\right)^{2}\\
 & -\frac{1}{8\pi G}(\partial\phi/\partial r)(\partial\phi/\partial r)-\frac{A^{2}}{mr^{2}}\varphi^{\dagger}\varphi-m\text{\ensuremath{\phi}}\varphi^{\dagger}\varphi,
\end{array}\label{eq:L4varphi}
\end{equation}
where $A$ is the separation constant or the angular momentum. 

Further, we use the Gaussian approximation for the semi-analytic calculations
which has been introduced by \citep{Gupta2017}, 

\begin{equation}
\varphi\left(t,r\right)=Ne^{-r^{2}/\left(2\sigma\left(t\right)\right)^{2}+ir^{2}\alpha\left(t\right)},\:\phi\left(t,\boldsymbol{x}\right)=\mu\left(t\right)e^{-r^{2}/\left(2\tau\left(t\right)\right)^{2}}\label{eq:varphi phi}
\end{equation}
where $N$ is the number density of the boson particles. The next
step is to integrate the Lagrangian Eq.(\ref{eq:L4varphi}), inputting
Eq.(\ref{eq:varphi phi}), on the entire space. This yields
\begin{align}
L= & -\frac{N\left(8A^{2}+6\hbar\sigma(t)^{4}\left(m\alpha'(t)+2\hbar\alpha(t)^{2}\right)+3\hbar^{2}\right)}{4m\sigma(t)^{2}}-\frac{\sqrt{2}gN^{2}}{8\pi^{3/2}\sigma(t)^{3}}-\frac{3\sqrt{\pi}\mu(t)^{2}\tau(t)}{16G}\\
 & \hspace{3cm}+\frac{2\sqrt{2}mN\mu(t)\sigma(t)\tau(t)^{4}\sqrt{\frac{2}{\sigma(t)^{2}}+\frac{1}{\tau(t)^{2}}}}{\left(\sigma(t)^{2}+2\tau(t)^{2}\right)^{2}},
\end{align}
from which we obtain the equations of motion for the variables $\sigma(t),\alpha\left(t\right),\mu\left(t\right),\mathrm{and}\tau\left(t\right).$
The equations for the last three variables $\alpha\left(t\right),\mu\left(t\right),\mathrm{and}\tau\left(t\right)$
are all algebraically solved. Putting them back to the above Lagrangian,
we obtain the reduced Lagrangian for the single variable $\sigma(t)$,
\begin{equation}
L_{\mathrm{eff}}=-\frac{N\left(8A^{2}+3\hbar^{2}\right)}{4m\sigma(t)^{2}}-\frac{gN^{2}}{4\sqrt{2}\pi^{3/2}\sigma(t)^{3}}+\frac{25\sqrt{\frac{5}{2\pi}}Gm^{2}N^{2}}{81\sigma(t)}+\frac{3}{4}mN\sigma'(t)^{2}.
\end{equation}
These ugly numerical coefficients come from the Gaussian approximation
and the values of them are of order one. Therefore, we simply analyze
the Lagrangian setting all the coefficients one. In particular, the
effective potential becomes 
\begin{equation}
V_{\mathrm{eff}}=\frac{gN^{2}}{\sigma(t)^{3}}-\frac{GM^{2}}{\sigma(t)}+\frac{J^{2}}{M\sigma(t)^{2}},\label{eq:Veff}
\end{equation}
where $M=mN$ and $J^{2}=N^{2}\left(2A^{2}+(3/4)\hbar^{2}\right)$. 

The last term of Eq.(\ref{eq:Veff}) comes from the angular momentum
and prevents the collapse of the system. In our previous study\citep{Morikawa2019},
we assumed the Axion DM and its inevitable self-interaction naturally
yields $g<0$. This tiny attractive self-interaction, the first term
in Eq.(\ref{eq:Veff}), can dominate the angular momentum for a massive
system and allows the SMBH formation. On the other hand in this article,
we try to avoid the angular momentum problem from a general point
of view not specifying DM species and not utilizing self-interaction.

Another necessary condition, on top of Eq.(\ref{eq:Kaup}), for the
BH formation is given by considering this effective potential $V_{\mathrm{eff}}$
Eq.(\ref{eq:Veff}) without the first term. A BH is formed when the
radius which gives the minimum of this potential is inside the corresponding
Schwarzschild radius. This condition yields
\begin{equation}
\mu\le1,\mathrm{where}\;\mu\equiv\frac{cJ}{GM^{2}},\label{eq:=00005Cmu}
\end{equation}
where $\mu$ is also known as the spin parameter of a BH. This parameter
$\mu$ can be understood as the ratio of the surface velocity to the
light velocity of a BH in the non-relativistic sense.

In order to achieve this condition $\mu\le1$ for SMBH formation,
we need to go back to the early stage of the Universe when $J$ was
small. This is because the present value of $\mu$ is too large $\mu\approx10^{4}$,
as we have observed in Eq.(\ref{eq:1}) and in Fig\ref{fig:The-angular-momentum}.
SMBH cannot be expected to form at the present stage even for the
coherent BEC-DM. This amount of angular momentum should be acquired
in the early stage of the evolution in the Universe. Therefore if
we go back to the early stage of the first acquisition of the angular
momentum, we have a chance to fulfill the condition $\mu\le1$, and
SMBH can be formed. 

\section{Tidal Torque Acquisition of Angular Momentum and BEC Collapse}

The angular momentum of any astronomical object will originate from
the tidal torque exerted in the early stage of the Universe. This
general process is precisely described by the tidal torque theory\citep{Peebles1971,Efstathiou1979}.
Suppose some portion of density fluctuation in the uniform background
starts to evolve, and deviates from the uniform cosmic expansion.
Then, this portion is necessarily exerted torque from its nearby region.
It acquires the angular momentum of the amount

\begin{equation}
\boldsymbol{J}(t)=\int_{a^{3}V}d^{3}x\rho\boldsymbol{x}\times\dot{\boldsymbol{x}}=\rho_{b}a^{5}\int_{V}d^{3}r(1+\delta)\boldsymbol{r}\times\dot{\boldsymbol{r}},\label{eq:Jt1}
\end{equation}
where the physical scale $\boldsymbol{x}$ is expressed by the comoving
scale $\boldsymbol{r}$ as $\boldsymbol{x}=a(t)\boldsymbol{r}$ and
$\rho_{b}$ is the background uniform density after removing the density
fluctuation factor $1+\delta$ \citep{Peebles1971,Efstathiou1979}.
Zel'dovich approximation\citep{Zel'dovich1970} $\boldsymbol{r}=\boldsymbol{q}-D(t)\boldsymbol{\nabla}\psi$
greatly reduces the above expression, where $\boldsymbol{\nabla}\psi$
is the peculiar gravitational potential and $D(t)$ is the fluctuation
growing factor. Further, the density enhancement $\delta$ can be
expressed by the Jacobian $(1+\delta)=\left|\partial\boldsymbol{q}/\partial\boldsymbol{\boldsymbol{r}}\right|$.
Then the standard $\Lambda$CDM cosmology predicts the evolution of
the angular momentum in the lowest order as 

\begin{equation}
\boldsymbol{J}(t)=\rho_{b}a_{0}^{3}a(t)\dot{D}(t)\int_{V}dq\boldsymbol{q}\times\boldsymbol{\nabla}\psi(\boldsymbol{q}).\label{eq:Jt2}
\end{equation}
Actually this theory is well verified by the numerical simulations\citep{Sugerman2000}.

We simply try the singular isothermal distribution for the above over-density
region in BEC-DM as 

\begin{equation}
\text{\ensuremath{\rho(t)}}=\frac{\beta(t)\text{\ensuremath{\rho_{0}}}}{\left(r/r_{0}\right)^{2}},\label{eq:rhot}
\end{equation}
where $r$ is the comoving length scale and $r_{0}$, $\rho_{0}$
are the constants although redundant. The time dependent factor $\beta(t)$
represents the development of the density enhancement in the early
Universe. We further assume the rigid rotation of this region and
the rotation velocity at the distance $r$ from the center is written
as 
\begin{equation}
v(t)=\alpha(t)\Omega r,\label{vt}
\end{equation}
where $\alpha(t)\Omega$ represents the increasing angular momentum
and $\Omega$ is a constant parameter. 

The density more increases inside but the angular momentum more increases
outside. Then, we expect a BH is formed at inner central region. Thus,
we consider the spherical region around the center with the radius
$r$. The total angular momentum inside the radius $r$ is given by
\begin{equation}
J(r)=\frac{4}{3}\pi\alpha(t)\beta(t)\text{\ensuremath{\rho_{0}}}r_{0}^{2}r^{3}\Omega,\label{eq:Jr}
\end{equation}
and the mass inside the radius $r$ by
\begin{equation}
M(r)=4\pi\beta(t)\text{\ensuremath{\rho_{0}}}r_{0}^{2}r.\label{eq:Mr}
\end{equation}

The necessary conditions for the BH formation are $\mu\le1$, Eq.(\ref{eq:=00005Cmu}),
and $M\geq M_{K}$, Eq.(\ref{eq:Kaup}). Using Eqs.(\ref{eq:Jr},\ref{eq:Mr}),
the central region inside the radius $r$ turns out to have the following
$\mu$ parameter,
\begin{equation}
\mu(r)\equiv\frac{cJ(r)}{GM(r)^{2}}=\frac{9\alpha(t)cr^{2}\Omega}{20\pi\beta(t)G\text{\ensuremath{\rho_{0}}}r_{0}^{2}r}.\label{eq:mur}
\end{equation}
If once this condition is satisfied, as well as $M(r)\geq M_{K}$,
then there is no preventing factor for this region to collapse to
a BH. 

According to the standard CDM model, the density enhancement evolves
as $\propto t^{2/3}$, and the factor $a(t)^{2}\dot{D}(t)$ in Eq.(\ref{eq:Jt2})
is proportional to time $t$. We use the time evolution of the angular
momentum and the mass enhancement numerically analyzed in the reference
\citep{Sugerman2000}. According to this analysis, the typical time
scale of the angular momentum acquisition is about 3 Giga years. Therefor,
we may set

\begin{equation}
J(t)\propto\alpha(t)\beta(t)\propto t,
\end{equation}

\begin{equation}
M(t)\propto\beta(t)=\left(\frac{t}{3G\text{ year }}\right)^{2/3}.
\end{equation}

Further, since the matter collapses to form BH without any resistance
if $\mu\le1$, the time needed to form BH can be estimated simply
by the free fall time of this matter, 

\begin{equation}
t_{\mathrm{ff}}=G^{-1/2}\left(\frac{M(r)}{\frac{4\pi}{3}r^{3}}\right)^{-1/2}.
\end{equation}
Therefore, the maximum mass BH will be formed at the free fall time
when the $\mu$ parameter becomes critical, we have the condition, 

\begin{equation}
t_{\mathrm{ff}}(t,r)=t\text{ and }\mu(t,r)=1.\label{eq:tfftmu1}
\end{equation}

After the collapse of the central region, the remaining outside part
will further acquire angular momentum and the SMBH formation may back
react to the outer region to increase the velocity dispersion there.
This outer part will be thus stabilized and forms DH surrounding the
central SMBH. Although this dynamical process is relevant to determine
the properties of the produced SMBH and the detail of DH, the necessary
calculation goes beyond the scope of this paper. This interesting
problem will be discussed in a separate report in the near future,
possibly including general relativistic numerical calculations. 

Thus, if we solve Eq.(\ref{eq:tfftmu1}), we can estimate the radius
$r$ inside of it forms the maximum mass BH at that time. Consequently,
the SMBH formation time scale is given by 
\begin{equation}
t_{\mathrm{BH}}=\text{ }\frac{2\sqrt{6\pi}}{c\Omega}\left(\frac{GM_{\mathrm{tot}}}{R_{\mathrm{tot}}}\right)^{1/2},
\end{equation}
and the region which collapses into SMBH has the radius 
\begin{equation}
r_{\mathrm{BH}}=\frac{6(3\pi)^{1/6}}{c^{4/3}t_{\mathrm{J}}^{1/3}\Omega^{4/3}}\left(\frac{GM_{\mathrm{tot}}}{R_{\mathrm{tot}}}\right)^{7/6}.
\end{equation}
Therefore the mass of the SMBH is given by 
\begin{equation}
M_{\mathrm{BH}}=\frac{12\sqrt{3\pi}}{c^{2}Gt_{J}\Omega^{2}}\left(\frac{GM_{\mathrm{tot}}}{R_{\mathrm{tot}}}\right)^{5/2}.
\end{equation}
Setting typical values, for example, 
\begin{equation}
M_{tot}=10^{12}\mathrm{M}_{\odot},R_{tot}=10\mathrm{kpc},\Omega=\frac{200}{R_{tot}}\frac{\mathrm{km}}{\mathrm{Sec}},t_{J}=3\text{ Giga year , }\label{eq:typical parameters}
\end{equation}
we have the SMBH formation time scale, the size which turns into SMBH,
and the resultant mass of SMBH as, 
\begin{equation}
\begin{array}{l}
t_{\mathrm{BH}}=0.9\times10^{6}\text{ years, }\\
r_{\mathrm{BH}}=20\mathrm{pc},\\
M_{\mathrm{BH}}=0.94\times10^{7}M_{\odot}\text{. }
\end{array}
\end{equation}
Thus the mass ratio of SMBH and dark halo (DH) $M_{DH}$ around it
is given by 
\begin{equation}
\frac{M_{BH}}{M_{DH}}\approx10^{-5}.\label{eq:MBH/MDH}
\end{equation}
Further, we can consider various sizes of the over-density regions
in BEC-DM. For this purpose, we tentatively assume a simple relation
for mass and size $M_{\text{tot }}\propto R_{\text{tot }}$for those
regions. Then the virial relation $v^{2}\sim GM/R$ yields almost
non-varying velocity and the assumption of the rigid rotation yields
$\Omega\propto v/R\propto v/M$. If we further assume that the typical
time scale of the object follows the evolution law for the linear
perturbation, then the mass ratio of SMBH and DH is given by the expression
\begin{equation}
\frac{M_{BH}}{M_{DH}}\approx10^{-5}\left(\frac{M_{tot}}{10^{12}\mathrm{M}_{\odot}}\right)^{-1/2},\label{eq:ratio1}
\end{equation}
as a function of the total mass $M_{tot}$.

However, these values strongly depend on the shape of the cluster
we assume as Eq.(\ref{eq:rhot}). If we have chosen a different density
profile
\begin{equation}
\text{\ensuremath{\rho}(t)}=\frac{\beta(t)\text{\ensuremath{\rho_{0}}}}{\left(r/r_{0}\right)^{5/2}},\label{eq:rhot2}
\end{equation}
with slightly steep configuration, then we have, under the same condition
Eq.(\ref{eq:typical parameters}), 
\begin{equation}
\begin{array}{l}
t_{\mathrm{BH}}=3.7\times10^{6}\text{ years, }\\
r_{\mathrm{BH}}=309\mathrm{pc},\\
M_{\mathrm{BH}}=2.02\times10^{9}M_{\odot}\text{. }
\end{array}
\end{equation}
The mass ratio of SMBH and DH is given, with the same assumption for
Eq.(\ref{eq:ratio1}), by 
\begin{equation}
\frac{M_{BH}}{M_{DH}}\approx2\times10^{-3}\left(\frac{M_{tot}}{10^{12}\mathrm{M}_{\odot}}\right)^{-1/2},\label{eq:ratio2}
\end{equation}
which is much larger than the previous isothermal case Eq.(\ref{eq:ratio1}).
On the other hand if we have chosen the density profile of shallower
configuration, then the formed SMBH mass and the ratio are much smaller.
The above estimates are based on bold analytic approximations. For
serious verification of our model and for the proper comparison with
observations, we need reliable numerical calculations based on general
relativity. 

\section{Observational Verification}

\begin{figure}
\includegraphics[width=12cm]{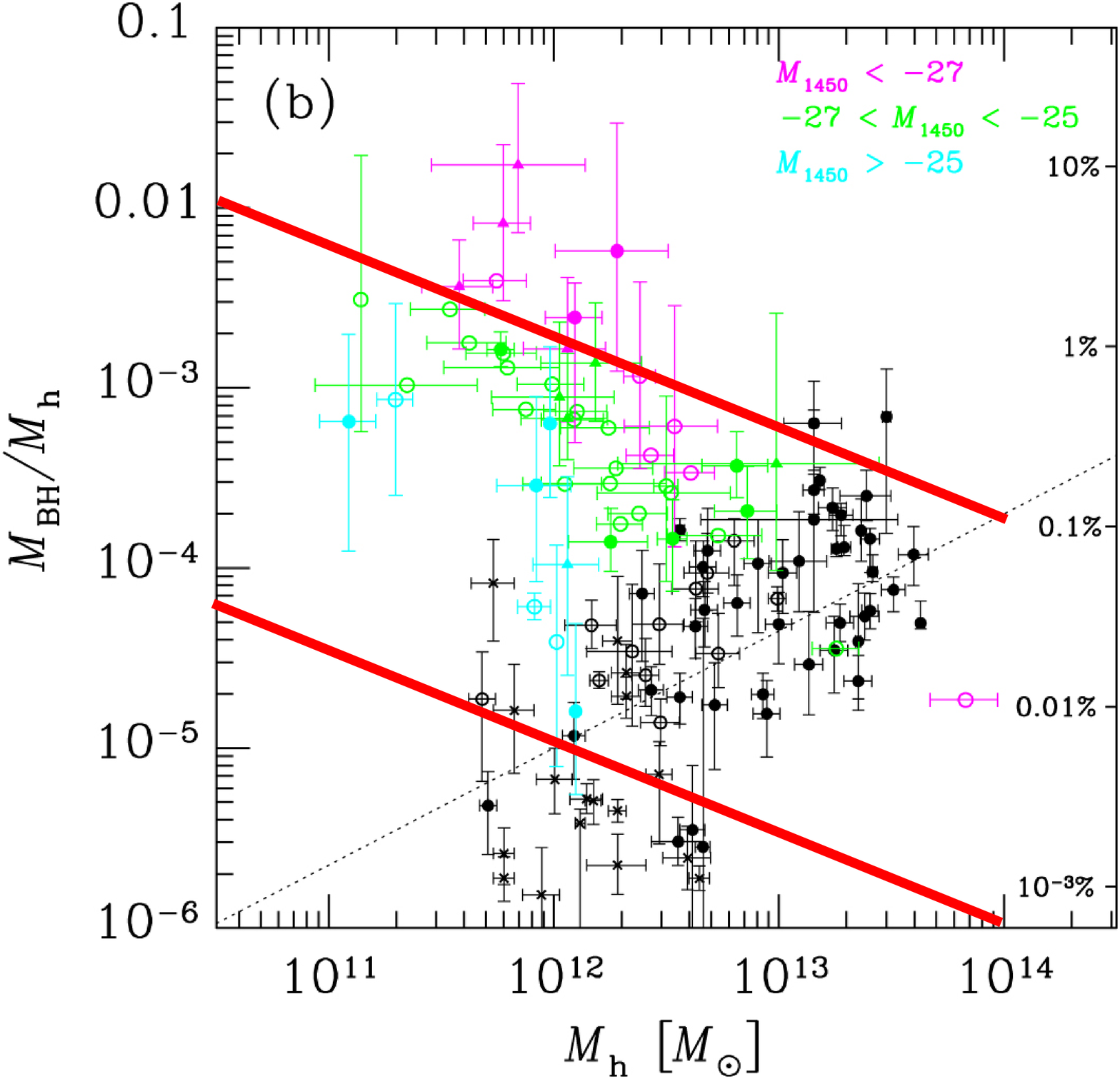}\caption{The mass ratio of SMBH and DH $M_{BH}/M_{DH}$ as a function of $M_{DH}$:
observational data and our estimates. On top of the observational
figure from \citep{Shimasaku2019} Fig3b, our estimates Eq.(\ref{eq:ratio1})(upper
red line) and Eq.(\ref{eq:ratio2})(lower red line) are plotted. $z\approx6-7$
objects are colored depending on $M_{1450}$(rest frame 1450\r{ }A
absolute magnitude: magenta, brighter than \textminus 27; green, \textminus 27
to \textminus 25; cyan, fainter than \textminus 25. Black symbols
are data of local galaxies quoted from \citep{Kormendy2013} in \citep{Shimasaku2019}:
ellipticals $\bullet$, classical bulges $\circ$, and pseudo bulges
$\times$. The detailed explanation is in \citep{Shimasaku2019}.
\label{fig:ratio}}

\end{figure}

We now turn our attention to the possible verification of our calculations.
Our main estimate has been the mass ratio of SMBH and DH. We examine
the observation of 49 QSO data around $z\approx6$ \citep{Shimasaku2019}
and the local data $z\approx0$ \citep{Kormendy2013} quoted in it.
The authors of \citep{Shimasaku2019} estimate $M_{DH}$ of a QSO
by measuring the CII velocity-width and assuming it as the circular
velocity of DH. Then they get the mass ratio of SMBH and DH $M_{BH}/M_{DH}$
as a function of $M_{DH}$. Their result Fig. 3 is quoted in our figure
\ref{fig:ratio}. On top of the observational data, we superposed
our results Eq.(\ref{eq:ratio1}) and Eq.(\ref{eq:ratio2}) by red
lines. We describe the verification of our results. 

All the observational data has great scatter and therefore there may
be no definite tendency in $M_{BH}/M_{DH}$. However, it is apparent
that the ratio $M_{BH}/M_{DH}$ of galaxies at high redshift $z\approx6-7$
is about two orders of magnitude larger than that of present galaxies
$z\approx0$. This fact may indicate that the evolution efficiency
of DH is far larger than that of SMBH. On the other hand, this result
may by caused simply by the bias effect to select luminous QSOs and
therefore $M_{BH}$ may be overestimated than the average. 

In the same way, our estimate of $M_{BH}/M_{DH}$ has large variation
mainly depending on the choice of the density profile. This is indicated
the wide separation of the two red lines in Fig.\ref{fig:ratio}.
However, if the density profile, as well as other values assumed,
were not actually unique, then we may need to observe more detail
and find possible correlations in the data for the verification of
our model. 

Since our estimate of the ratio $M_{BH}/M_{DH}$ is for the SMBH-DH
system just formed in the high redshift era, our result should be
compared more with objects at redshift $z\approx6-7$ rather than
that at $z\approx0$. Further, our detailed dynamical analysis for
the evolution of SMBH and DH may reveal the reason for the large reduction
of the rate from $z\approx6-7$ to $z\approx0$. 

Overall comparison of our estimate and the observational data are
not so different from each other and it does not exclude our model.
In the comparison, the slope $-1/2$ for the ratio should not be taken
seriously since the slope $-1/2$ is the consequence of our naive
assumption $M_{\text{tot }}\propto R_{\text{tot }}$in our calculations.
The validity of this assumption as well as others should be checked
in the next step of our research. 

The authors of \citep{Shimasaku2019} quote several other observational
results for the ratio $M_{BH}/M_{DH}$ in different redshift $z\approx2-3,4-5$
(Fig 5. in \citep{Shimasaku2019}). They also show large data range
$10^{-5}\le M_{BH}/M_{DH}\le10^{-2}$. Our estimates Eq.(\ref{eq:ratio1},\ref{eq:ratio2})
are not excluded by the comparison with these data. We would like
to explore our calculations to find other possible correlations than
$M_{BH}/M_{DH}$ for further verification in the future. 

\section{Conclusions and Discussions }

We have discussed the origin of SMBH formed from the BEC-DM solving
angular momentum problem. This scenario has been needed to guarantee
the very early formation and the common existence of SMBH suggested
from recent observations. We first examined the prevailing angular
momentum associated with all astronomical objects. The angular momentum
turns out to follow the scaling law in the wide mass range and the
amount of it is about $10^{4}$ times larger than the critical value
for black hole formation. However, we obtained a good motivation that
the SMBH may be formed from DM. We applied the self-gravitating-turbulence
picture and tentatively estimated the time scale for the structure
formation at each mass scale. For example, the formation of SMBH of
mass $10^{9}M_{\odot}$ needs $10^{9}$years, which does not meet
observations. Furthermore, these arguments only apply to the steady
stationary processes and therefore cannot be applied to much larger
scales where the dynamical evolution effect dominates. 

Therefore we proceed to consider the dynamical formation of SMBH based
on BEC-DM which is described by non-relativistic approximation (GP
and Poisson equations). After a brief introduction of the BEC-DM scenario
of SMBH formation, we applied the variable separation method and the
Gaussian approximation to obtain the effective action, which yields
the approximate BH formation condition within the non-relativistic
calculations. 

Then estimating the tidal torque acquisition of the angular momentum
and setting appropriate density profiles, we obtain the time scale
of SMBH formation and the collapsing scale as well as the maximum
mass of SMBH. Then simply assuming the relation between the mass and
the length of the initial DM cluster, we obtain the mass ratio of
the SMBH formed and the remaining DH as a function of the total initial
mass. The range of the estimated mass ratio widely varies, $10^{-5\sim-3}$at
the mass scale $10^{12}M_{\odot}$, for different density profiles
of the initial DM cluster. 

Finally, these estimates were compared with the recent observations
of the SMBH/DH ratio. We compare our results with the high redshift
data and the local data. It turns out that our estimate is not excluded
by the observations although the observational data has a large scatter
and therefore allows a wide range of predictions. Moreover, the mass
ratio is reported to have a huge evolution effect and reduces 1-2
orders of magnitude from $z\approx6$ to $z\approx0$.

We found that only a small central portion of the DM cloud can form
SMBH and the rest of the cluster simply becomes DH in the very early
stage when the cluster is formed and the tidal torque accumulates
its angular momentum. Thus the angular momentum controls the separation
of the original DM cluster into the central SMBH and the surrounding
DH. If there were no angular momentum, then the whole DM cluster collapses
to form an extremely huge BH. Subsequent galaxy, if formed around
it, would be quite peculiar and fully different from what we observe
today. Thus our model, angular momentum controlled SMBH/DH formation
from BEC-DM, guarantees the rapid formation and the common existence
of SMBH at the center of DH, in which many stars and a galaxy are
going to evolve. However, we have used several analytic approximations
without fully examining their validity. Actually, the assumption of
the density profile has drastically changed the mass of the formed
SMBH. It is apparent that we need much elaboration of our model in
the future.

We have proposed the scenario of the early formation of SMBH from
BEC-DM avoiding the angular momentum problem. The present calculation
can be improved, the scenario can be much precise, and the detailed
comparison with observations can be possible if we further consider
the following subjects, including general relativity and numerical
methods. Although we have assumed the BH formation proceeds only
at the center of the cluster, many smaller BHs may be formed in the
other regions of the cluster as well if we assume the quiet acquisition
of angular momentum. However, when the central SMBH is formed, a strong
shock wave may heat up the surrounding DH region and disturb the formation
of those smaller BHs. A numerical calculation of this complicated
process would be interesting to get the mass distribution function
of BHs.  Further, the first formed SMBH in our scenario determines
the center of the galaxy. It may promote star formation by the shock
wave associated with the BH jet. Eventually, the bulge mass and the
SMBH mass may establish the observed universal correlation. This process
has been partially analyzed before \citep{Morikawa2021}, but has
not been well analyzed so far. We hope to report all of these calculation
results in our subsequent publications soon. 

\section*{Acknowledgments}

We wish to acknowledge the fruitful discussions with all the members
of the astrophysics group of Ochanomizu University in particular Sakura
Takahashi.

\end{document}